\documentclass[aps,pre,twocolumn]{revtex4-1}
\usepackage{amssymb}
\usepackage{amsmath,bm}
\usepackage{graphicx,color}
\usepackage{bbm}
\usepackage{multirow}
\newcommand{\B}[1]{{\bm{#1}}}

\begin{document}

\title{Robustness of Density of Low Frequency States in Amorphous Solids}
\author{Prasenjit Das$^1$, H. George E. Hentschel$^{1,2}$, Edan Lerner$^3$ and Itamar Procaccia$^{1,4}$}
\affiliation{$^1$Department of Chemical Physics, The Weizmann Institute of Science, Rehovot 76100, Israel.\\
$^2$ Dept. of Physics, Emory University, Atlanta Ga. 30322, USA.\\ $^3$ Institute for Theoretical Physics,
University of Amsterdam, 1098 XH Amsterdam, the Netherlands. \\$^4$  Center for OPTical IMagery Analysis and Learning, Northwestern Polytechnical University, Xi'an, 710072 China.  }

\begin{abstract}
Low frequency quasi-localized modes of amorphous glasses appear to exhibit universal density of states, depending on the frequencies as $D(\omega) \sim \omega^4$. To date various models of glass formers with short range binary interaction, and network glasses with both binary and ternary interactions, were shown to conform with this law. In this paper we examine granular amorphous solids with long-range electrostatic
interactions, and find that they exhibit the same law. To rationalize this wide universality class we return to a model proposed by Gurevich, Parshin and Schober (GPS) and analyze its 
predictions for interaction laws with varying spatial decay, exploring this wider than expected universality class. Numerical and analytic results are provided for both
the actual system with long range interaction and for the GPS model. 
\end{abstract}

\maketitle

\section{Introduction}

It had been known for more than thirty years now \cite{83KKI,87IKP,91BGGS,03GC,03GPS,07PSG,14SBG}  that low-frequency vibrational modes in amorphous glassy systems are expected to present
a density of states $D(\omega)$ with a universal dependence on the frequency $\omega$, i.e.
\begin{equation}\label{dof}
D(\omega)\sim \omega^4 \ .
\end{equation}
The direct verification of this prediction is not always straightforward, since the modes which are expected to exhibit this universal scaling are quasi-localized modes (QLM) that
in large systems hybridize strongly with low frequency delocalized elastic extended modes. The latter modes have another universal form; their density of states depends on frequency like $\omega^{d-1}$ where $d$ is the spatial dimension. To observe the universal law Eq.~(\ref{dof}) one needs to disentangle these different types of modes. Numerical simulations are ideal for this purpose since by necessity they are
limited to relatively small systems in which the (Debye) delocalized modes have a lower cutoff, exposing cleanly the QLM with their
universal density of states Eq.~(\ref{dof}). In fact, the lowest Debye mode is expected to have a frequency of the order of $2\pi c/L$ where $c$
is the speed of sound and $L$ is the system size. Thus in smaller systems the lower cutoff of the Debye modes is pushed up. In the recent literature there were a number of direct verifications of this law,
using numerical simulations of glass formers with binary interactions \cite{16LDB,15BMPP,shimada2018spatial,Moriel2019,angelani2018probing,17MSI,18KBL}, and also more recently in models of silica glass
with binary and ternary interactions \cite{20BGMPS,20GRKPVBL}. 

The aim of this paper is to examine how {\em long-ranged interactions} effect the density of states. To this aim
we return to a model studied recently of charged disks and spheres in two and three dimensions respectively \cite{20DHP}. On the face of it 
the existence of long-ranged interactions could introduce strong deviation from the universal law (\ref{dof}). To our surprise it
turned out that the scaling law (\ref{dof}) is very robust, and the addition of long-ranged interactions did not alter it.
To understand this we return to the very interesting theory offered by Gurevich, Parshin and Schober (GPS) which over twenty years expounded
their understanding of the origin of the scaling law (\ref{dof}). In the third section of this paper we paraphrase their derivation,
extending it to interaction laws that were not treated by these authors. We make a special attempt to stress the main assumptions and
approximations that underlie the proposed universality. In particular we seek interactions and parameters that should, according to the
theoretical analysis, lead to a failure of Eq.~(\ref{dof}) in order to further clarify when and where the universality is expected to hold.
We find that in fact the law Eq.~(\ref{dof}) is very robust, even when the conditions for the existence of the theory are not available. 
In some sense the universality of Eq.~(\ref{dof}) is broader than one could anticipate.  

The structure of this paper is as follows: in Sec. \ref{model} we discuss the model of charged disks or spheres, and present
the numerical results for the density of states. In Sec. \ref{theory} we offer a review of the theory of Refs. \cite{91BGGS,03GC,03GPS,07PSG} 
with strong emphasis on the assumptions and approximations made. In Sec.~\ref{GPS} we examine the GPS model with various laws
of interaction, with a stringent test on the applicability of the assumptions in the theory. A summary and conclusions are offered
in Sec.~\ref{summary}.

\section{Charged Granular Model}
\label{model}
\subsection{Model Properties}
\label{properties}
To examine the density of states in charged compressed granular media we study a model consisting of a 50-50 mixture of $N$ frictionless two-dimensional disks or three-dimensional spheres
 with diameters $R_1=1.0$ and $R_2=1.4$ respectively. Below all the lengths are measured in the unit of $R_1$. Half of the smaller particles are positively charged at the center of mass with a charge $+q$ and the other half are 
 negatively charged with a charge $-q$. The same is true for the large ones. The particles are placed randomly inside a two-dimensional or three-dimensional box such that there is no overlap between two particles.  Molecular dynamics is then used to equilibrate the system.
During this equilibration one adds a damping term $-\kappa \dot {\B r_i}$ to each of the 
 equations of motion. Once equilibrated,  the system is compressed in small steps to achieve a required packing fraction $\phi>\phi_J$,
 with the above mentioned damping to equilibrate the system after each step. The value of 
$\phi_J \approx 0.843$ in 2d and 0.639 in 3d, being the jamming packing fraction at zero temperature (for uncharged systems). The result of this procedure is an equilibrated amorphous solid that is charge-neutral. 
 The simulation presented below employs periodic boundary conditions, the total number of particles is $N=600, 1000, 2000$ and 8000
 in 2d and $N=$3600, 4500 and 5500 in 3d. In all cases $\phi=0.90$ in 2d and $\phi=0.67$ in 3d.

The short-ranged forces between two overlapping disks are Hertzian-elastic. The potential for these forces is given by \cite{01SEGHLP}:
\begin{equation}
	\label{eqn1}
	\varPhi_{elas}(r_{ij}) = \frac{2}{5}K_n\sqrt{R_{eff}}(R_{ij}-r_{ij})^{5/2}.
\end{equation}
Here, $K_n=20,000$ is an elastic
 constant. Denoting the centers of mass of the $i$th and $j$th disk as $\B r_i$ and $\B r_j$ then $r_{ij} = |\B r_i-\B r_j|$, $R_{ij} =  (R_i + R_j)/2$ and $R_{eff} = 0.5R_i R_j/(R_i+R_j)$. 

Apart from the elastic force, grains interact via long-ranged electrostatic forces. If $q_i$ and $q_j$ are the charges in the $i$th and $j$th grains, the electrostatic interaction potential is given, in Gaussian units,  by
\begin{align}
\label{eqn2}
\tilde V_{\rm elec}(r_{ij}) = \frac{q_iq_j}{ r_{ij}},
\end{align}
In our simulation we use units of charge such that $q_i=\pm 1$.
The electrostatic interaction is of course long-ranged. However, it has been shown \cite{06FG,07CBHIK} that in an amorphous mixture of randomly distributed charged grains, one can use the damped-truncated Coulomb potential as given by
\begin{align}
\label{eqn3}
V_{\rm elec}(r_{ij}) =  q_iq_j\left[ \frac{erfc(\alpha r_{ij})}{r_{ij}} - \frac{erfc(\alpha R_c)}{R_c} \right], r_{ij}\leq R_c \ ,
\end{align}
with $R_c$ being the cutoff scale of electric interaction, with $R_c=7.5$ in three-dimensions and 12.5 in two-dimensions.
Here $erfc(x)$ is the complementary error function, $\alpha=0.15$ is the damping factor of the electrostatic interaction due to screening. Below we employ the Hessian matrix, which is the second derivative of the potential with respect to coordinates. We therefore smooth out $V_{elec}$ at $r=R_c$ to have four derivatives when $V_{\rm elec}$ goes to zero at $r=R_c$. To this aim we use the following form
\begin{align}
	\label{eqn4}
	\varPhi_{\rm elec}(r_{ij}) = V_{\rm elec}(r_{ij}) - \sum_{n=1}^{4}\frac{(r_{ij} - R_c)^n}{n!}\left.\frac{d^nV_{\rm elec}}{dr_{ij}^n}\right|_{r_{ij}=R_c}.
\end{align}
At this point we should add that we have checked the sensitivity of our numerical results to the choice of $R_c$. The
qualitative nature of the results did not change when $R_c$ was chosen larger, but the numerical effort was increased by
much. 

The total binary potential $\varPhi(r_{ij})$ is therefore
\begin{equation}
\varPhi(r_{ij}) \equiv	\varPhi_{\rm elas}(r_{ij})+ \varPhi_{\rm elec}(r_{ij}) \ .
\label{defphi}
\end{equation}

Finally, the Hessian matrix is given by:
\begin{equation}
H_{ij}^{\alpha\beta}\!=\!-\Big(\frac{\partial^2\Phi(r_{ij})}{\partial r_{ij}^2} \! -\! \frac{1}{r_{ij}}\frac{\partial\Phi(r_{ij})}{\partial r_{ij}} \Big)n_{ij}^\alpha n_{ij}^\beta
- \frac{\delta_{\alpha\beta}}{r_{ij}}\frac{\partial\Phi(r_{ij})}{\partial r_{ij}} \ ,
\end{equation}
where $n_{ij}^\alpha=(r_j^\alpha - r_i^\alpha)/r_{ij}$. The diagonal elements of the Hessian matrix read
\begin{equation}
H_{ii}^{\alpha \beta} =- \sum_{\ell\ne i} H_{i\ell}^{\alpha \beta} \ .
\end{equation}
The Hessian matrix, being real and symmetric, has real eigenvalues. Besides Goldstone modes associated
with continuous translational symmetries that yield two (three) zero eigenvalues in two (three) dimensions, all the other eigenvalues are positive as long as the system is
mechanically stable. Every eigenvalue $\lambda_i$ of the Hessian matrix is associate with a frequency 
\begin{equation}
\omega_i \equiv \sqrt{\lambda_i} \ .
\end{equation}
The density of states refers to the probability distribution function $D(\omega)$ of these frequencies in the limit $N\to \infty$.
\subsection{Numerical computation of the density of states}
\label{dos}

\begin{figure}
	\includegraphics[width=0.35\textwidth]{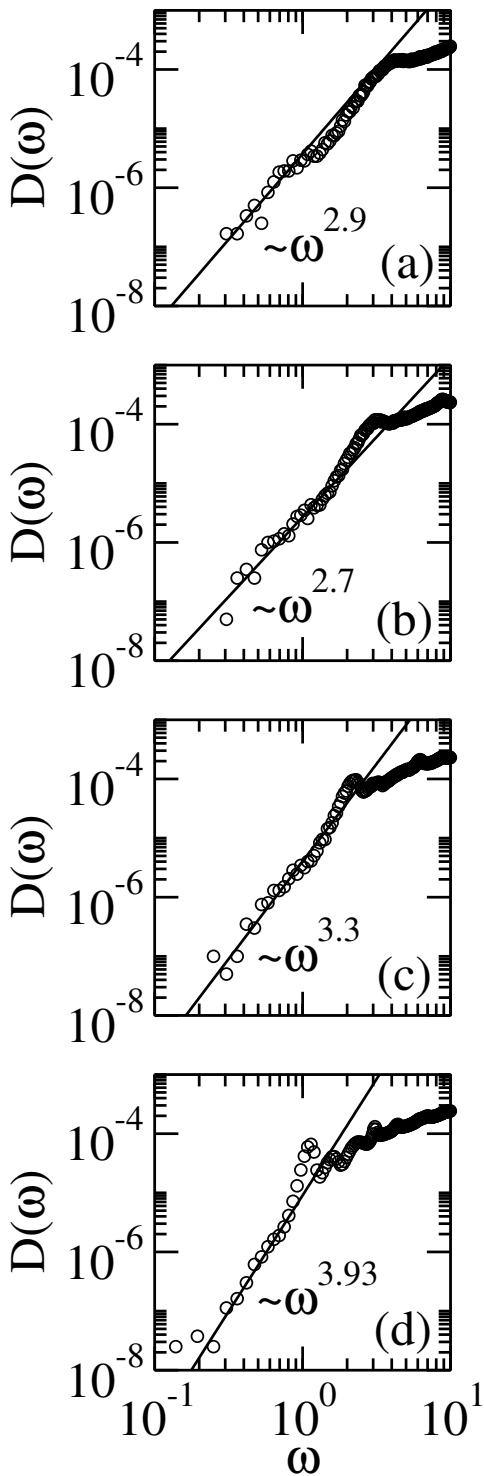}
	\caption{ The low (small frequency) tail of the density of states of the two-dimensional model of charged disks discussed in 
		Sec.~\ref{model} for four different system sizes, i.e. $N=600, 1000, 2000$ and 8000 in panels a-d respectively. In all cases
		we employed 10000 independent configurations. The lines are best fits
		to the small frequency tail of the DOS. 
		We conclude from this data that the density of states converges to the universal law Eq.~(\ref{dof}).}
		\label{2d}
		\end{figure}

\begin{figure}
	\includegraphics[width=0.35\textwidth]{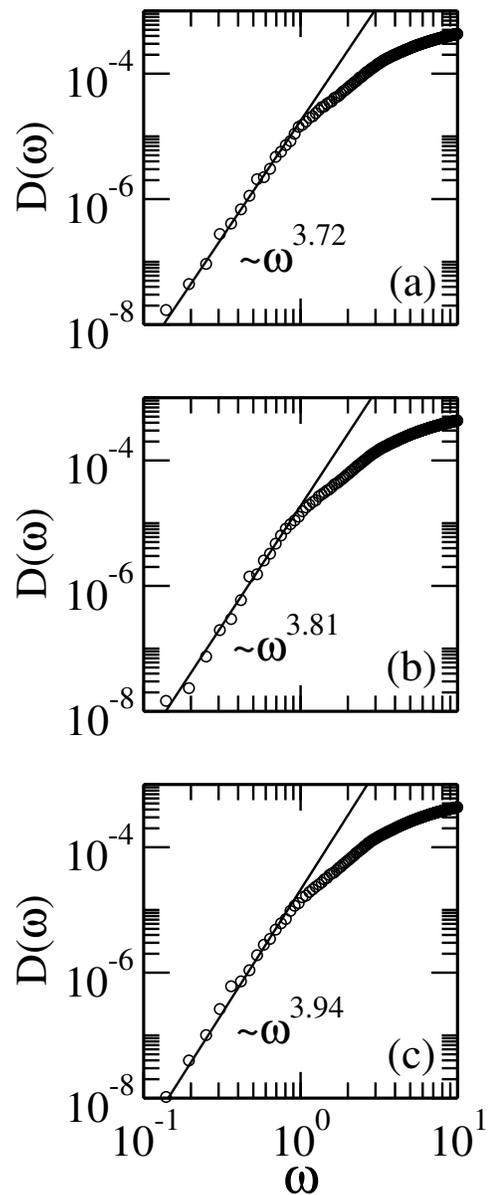}
	\caption{ The low (small frequency) tail of the density of states of the three-dimensional model of charged spheres discussed in 
		Sec.~\ref{model} for three different system sizes, i.e. $N=3600, 4500$ and 5500 in panels a-c respectively. In all cases
		we employed 5000 independent configurations.The lines are best fits
		to the small frequency tail of the DOS.
		We conclude from this data that the density of states converges to the universal law Eq.~(\ref{dof}).}
		\label{3d}
	\end{figure}

The density of states of our charge-neutral system is computed in both two and three dimensions for various system sizes.
It is important to ascertain the convergence of the density of states since one observes strong finite size effects, besides
the obvious remark that a ``density" exists only in the limit $N\to \infty$ \cite{20Ler}. In Figs.~\ref{2d} and \ref{3d} we present
the low-end (small frequency) regime of the density of states of the the model discussed above for four (three) system sizes in
two (three) dimensions. 

\begin{figure}
	\includegraphics[width=0.35\textwidth]{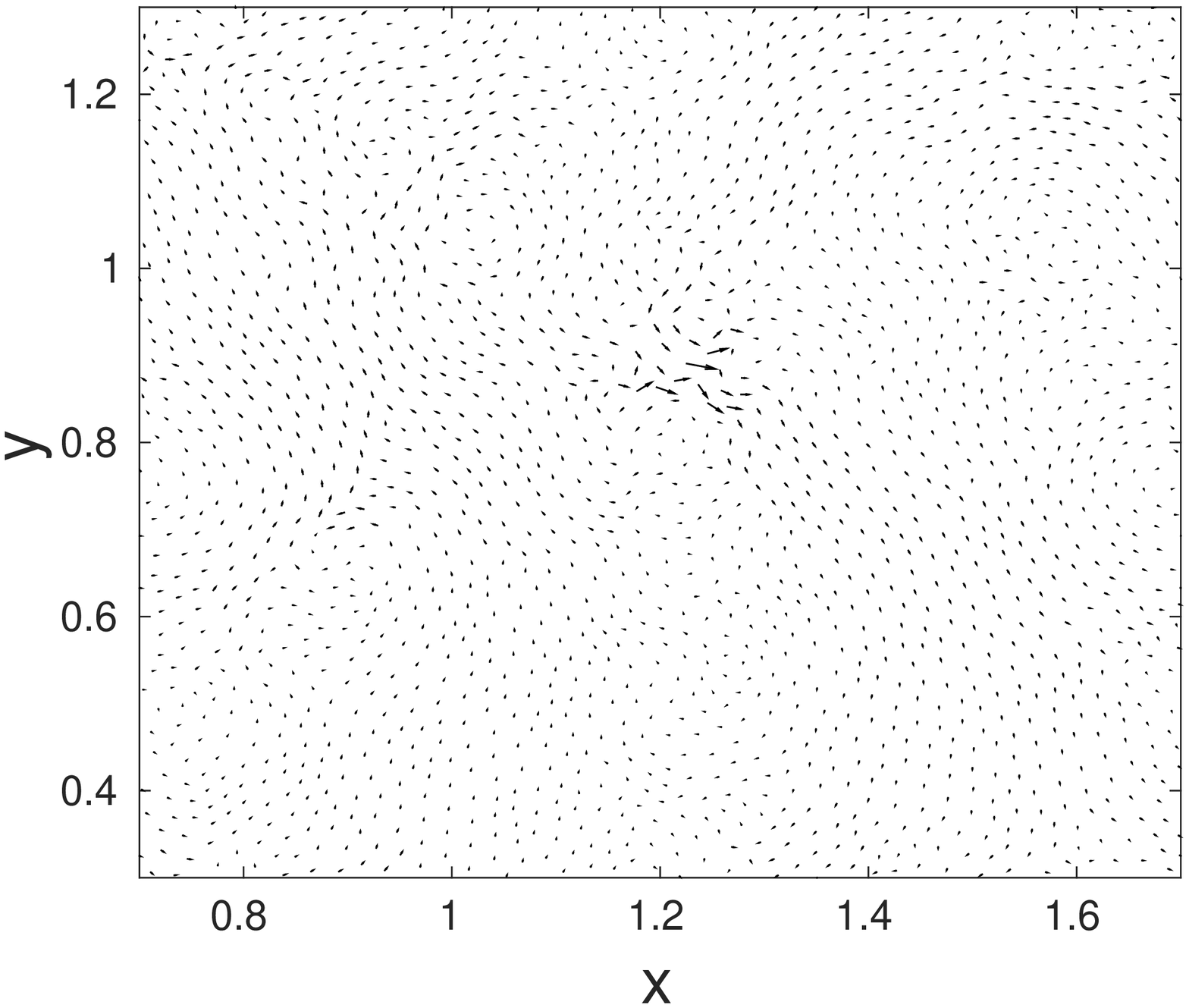}
	\caption{A representative example of an eigenmode in 2-dimensions of the Hessian matrix whose frequency 
		lies in the range of the universal $\omega^4$ regime. The participation ratio of this mode is 5.9$\times 10^{-2}$. It is obviously quasi-localized.}
	\label{mode2}
\end{figure}

\begin{figure}
	\includegraphics[width=0.35\textwidth]{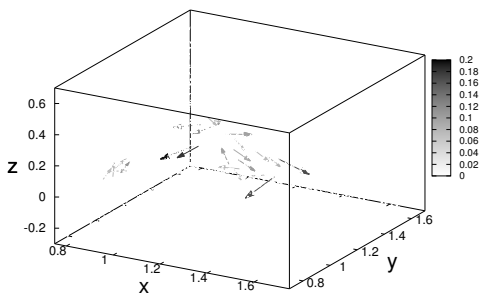}
	\caption{A representative example of an eigenmode in 3-dimensions of the Hessian matrix whose frequency 
		lies in the range of the universal $\omega^4$ regime. The participation ratio of this mode is 3$\times 10^{-2}$. It is obviously quasi-localized.}
	\label{mode3}
\end{figure}

The conclusion that we draw is quite obvious, i.e. that the universality class of Eq.~(\ref{dof}) includes the
present system despite the long range interactions, in both two and three dimensions. As seen in other cases, there is a system size 
dependence, with the slope approaching 4 when the system size increases \cite{20Ler}. The price is that for larger systems Debye modes and hybridization
penetrate lower frequencies \cite{16LDB}, shortening the regime of the universal power law,  as can be seen in Fig.~ \ref{3d}. 
It is important to realize that the
modes that participate in the scaling law  Eq.~(\ref{dof}) are all quasi-localized modes, rather than extended Debye modes.
To exemplify this we present in Figs.~\ref{mode2} and \ref{mode3} representative eigenfunctions whose eigenvalues are in the range
of the universal scaling law. These are obviously not extended modes, as one can also check by evaluating their participation
ratio. 
To rationalize the wideness
of the universality class we return now to the theory proposed in Refs. \cite{91BGGS,03GC,03GPS,07PSG}.
\section{The universality of the density of states}
\label{theory}

In the series of papers, Refs. \cite{91BGGS,03GC,03GPS,07PSG} and related ones, the authors attempted, over
a span of more than thirty years, to establish the universality of Eq.~(\ref{dof}) for amorphous glassy solids. The authors
 suggested that this universality is due to a vibrational instability of the spectrum of weakly interacting quasi-localized harmonic modes that is also responsible for the maximum in the function $D(\omega )/\omega^2$ in glasses, known as the the boson peak. Here we extend the derivation to different laws of interaction, using the opportunity to stress verbatim what are the essential assumption and approximation made to reach the law Eq.~(\ref{dof}). 

The theory is based on the notion that the normal modes of the Hessian can be represented as a collection of interacting oscillators. Once this is accepted, the derivation has two essential parts.  First, a complete reconstruction of the ``bare" vibrational DOS below some frequency $\omega_c$ takes place, proportional to the strength of the elastic interaction between these oscillators.  
This first reconstruction is independent of the bare DOS $g_0(\omega)$, subject to conditions on $g_0$ that are exposed below. This first
reconstruction leads to a density of states that is linear in $\omega$
\begin{equation}
\label{eq1}
g_1(\omega ) = C \omega ,
\end{equation}
and is limited by anharmonicity. The DOS of the new harmonic modes is independent of
the actual value of the anharmonicity for $\omega < \omega_c$. This first reconstruction is discussed here.

\subsection{First reconstruction}
\label{first}

This vibrational instability is a rather general phenomenon and occurs in any system of bilinearly coupled harmonic oscillators. It can be considered in a purely harmonic approximation.  We emphasize here that typically there are only a small number of low frequency oscillators with degrees of freedom $x_1$ with $\omega_1 \ll \omega_c$ (to be estimated below) surrounded by a large number of high frequency modes such as $x_2$ with $\omega_2 \gg \omega_c$, with typically $\omega_2 \approx \omega_{debye}$. 
For example consider a simplified model consisting of only two oscillators $x_1$ and $x_2$ with effective masses $M_1$ and $M_2$ and harmonic frequencies $\omega_1\ll \omega_c$ and $\omega_2\gg \omega_c$,
interacting via an interaction $I_{12}$ due to an elastic field. This leads to  a potential
\begin{equation}
\label{pot}
U(x_1,x_2) = M_1 \omega_1^2 x_1^2/2 + M_2 \omega_2^2 x_2^2/2 + I_{12} x_1 x_2.
\end{equation}
The interaction $I_{12}$ will renormalize the frequencies to ${\hat \omega^2}_{1,2}$ where
\begin{equation}
\label{eq2}
{\hat \omega^2}_{1,2} = \frac{\omega_1^2+\omega_2^2}{2} \mp \sqrt{[(\frac{\omega_1^2-\omega_2^2}{2})^2+ \frac{I_{12}^2}{M_1 M_2}]}. 
\end{equation}
From Eq.~(\ref{eq2}) we see that the smaller frequency ${\hat \omega}_1$  becomes zero and an instability occurs when $I > I_c$, where
\begin{equation}
\label{Ic}
I_c = \omega_1 \omega_2 \sqrt{M_1 M_2}.
\end{equation}
From Eq.~(\ref{Ic}) we can now estimate $\omega_c$ as follows. It is defined as the maximum frequency that can be destabalized by the interaction $I$. Using $M_1 \sim M_2 \sim M$, $\omega_1 \approx \omega_c$, and $\omega_2 \sim \omega_{debye}$ in Eq.~(\ref{Ic}) we find
\begin{equation}
\label{eq33}
\omega_c \approx \frac{ I_c}{ M \omega_{debye}} \ll \omega_{debye}.
\end{equation}
For $I_{12}=I < I_c$ the smaller frequency approaches zero as ${\hat \omega^2}_1 = \omega_1^2 [1-(I/I_c)^2]$, or
\begin{equation}
\label{hatom}
|{\hat \omega}_1| = \omega_1 \sqrt{ [1-(I/I_c)^2]}.
\end{equation}

Consider now a stable collection of oscillators with a density of such low frequency oscillators embedded in a sea of high energy oscillators.  In other words, 
we assume that if there were unstable oscillators with $I>I_c$ these were already reorganized when the system was equilibrated by energy minimization. 
Using Eq.~(\ref{hatom}) we can now write down a general expression for the reconstructed DOS of an initial DOS $g_0(\omega_1)$ due to a quenched distribution $P(I)$ of forces $I$ as
\begin{equation}
\label{eq5}
g_1(\omega ) = \int d\omega_1 g_0(\omega_1) \int_0^{I_c} dI P(I) \delta(\omega - \omega_1 \sqrt{ [1-(I/I_c)^2]}).
\end{equation}
To proceed we will {\em assume} that the distribution $P(I)$ is smooth and non singular at the vicinity of $I=I_c$.
Performing the $I$ integration over the delta function we end up with the expression
\begin{equation}
\label{eq6}
g_1(\omega ) = I_c \omega \int d\omega_1 g_0(\omega_1) \frac{P(I_c \sqrt{[1-\omega^2/\omega_1^2]})}{\omega_1^2 \sqrt{[1-\omega^2/\omega_1^2]}}.
\end{equation}
If we now take the $\omega \rightarrow 0$ of Eq.~(\ref{eq6}) we find
\begin{equation}
\label{g1}
g_1(\omega ) = I_c P(I_c) \omega \int d\omega_1 \frac{g_0(\omega_1)}{\omega_1^2}.
\end{equation}
Looking at Eq.~(\ref{g1}) we see that 
\begin{equation}
g_1(\omega ) \rightarrow C \omega\quad \text{as}~ \omega \rightarrow 0 \ ,
\label{linear}
\end{equation}
 with 
 \begin{equation}
C = I_c P(I_c) \int d\omega_1 \frac{g_0(\omega_1)}{\omega_1^2}\ ,
\label{C}
\end{equation}
{\em provided the integral converges}. 

It should be stressed here that it is not obvious what is the functional form $g_0$ for a given example of amorphous solid. If we take $g_0(\omega_0) \approx g_{debye}(\omega_0) \sim \omega_0^{d-1}$ we see that this integral converges for $d>2$, but for $d=2$ there would exist logarithmic corrections to the integral. There is no reason however to assume that the bare density of the QLM's is the same as the Debye
modes which are extended. So we need here to continue by faith, assuming that the integral converges. Then there is a universal behavior $g_1(\omega ) \sim \omega$ though $C$ is not universal. We will show below (cf. Sec.~\ref{GPS}) a case where this first reconstruction does not apply because $g_0$ is chosen such that the integral in Eq.~(\ref{C}) does not converge.

Another source of worry can arise if the distribution function $P(I)$ were {\em not} benign in the vicinity of $I_c$. For example imagine that in the vicinity of $I_c$ $P(I)\propto (I_c-I)^\beta$.
In that case we could return to the analysis described above and find that 
\begin{equation}
g_1(\omega) \propto \omega^{1+2\beta} \ , \quad \text{if} ~P(I)\propto (I_c-I)^\beta \ .
\end{equation}
Since we have no control at this point on the pdf of the interactions $I_{ij}$ we should bare in mind that this, and other possibilities that are not treated here explicitly can
bust the first reconstruction to Eq.(\ref{linear}).
\subsection{Second Reconstruction}
\label{second}

The first reconstruction discussed above pertains to frequencies smaller than some cutoff frequency $\omega_c$.
This second reconstruction mainly effects the DOS at frequencies $\omega \ll \omega_b \ll \omega_c$ where $D(\omega) \sim \omega^4$.

The second reconstruction comes about due to a further interaction between the oscillators which we have not taken into account so far. The low- frequency QLMs,  displaced from their equilibrium positions, create random quenched static forces $f_i$ on each QLM. The force $f_i$ exerted on the $i$th
oscillator by the other $j$ oscillators  is 
\begin{equation}
f_i = \sum_j J_{ij} x_{j0} \ .
\label{fi}
\end{equation} These forces have some distribution $P(f)$ . In the purely harmonic case, these linear forces would not
affect the frequencies. Anharmonicity, however, renormalizes the low frequency part of the spectrum. 

Consider an anharmonic oscillator under the action of a random static force $f$ given by the potential 
\begin{equation}
V(x) = M\omega_1^2 x^2/2 + A x^4/4 - f x \ .
\end{equation}
The force $f$ then shifts the equilibrium position from $x=0$ to $x=x_0$, given by
\begin{equation}
\label{shift}
Ax_0^3 + M \omega_1^2 x_0 - f =0 \ ,
\end{equation}
where the oscillator now has a new harmonic frequency given by $M\omega_{new}^2 = d^2V(x)/dx^2 |_{x=x_0}$ or
\begin{equation}
\label{eq3}
\omega_{new}^2 = \omega_1^2 + 3 A x_0^2/M .
\end{equation}
Now given $g_1(\omega)$ by Eq.~(\ref{eq1}) and a  distribution function for the random frequencies $P ( f )$, then the renormalized DOS is given by
\begin{equation}
\label{Domega}
D(\omega ) = \int_0^{\infty}g_1(\omega_1) d\omega_1 \int_{-\infty}^{\infty} df P(f) \delta(\omega - \omega_{new}) \ .
\end{equation}
 We note that because of the Dirac delta function in the second integral only values of the force $f$ where $\omega - \omega_{new}(\omega_1, f) =0$ will contribute to the integral. We can find an expression for $\omega_{new} (\omega_1, f)$ as follows. First from Eq.~(\ref{shift}) for small $f$ we find $x_0 \approx f/(M \omega_1^2 )$. If we substitute this form into Eq.~(\ref{eq3}) we find
\begin{equation}
\label{omeganew}
\omega_{new} (\omega_1, f) = \sqrt{\omega_1^2 + \frac{ 3 A f^2}{M^3 \omega_1^4}} > \omega_1 .
\end{equation}
and write Eq.~(\ref{Domega}) more explicitly as
\begin{equation}
\label{yes}
D(\omega ) = \int_0^{\omega}g_1(\omega_1) d\omega_1 \int_{-\infty}^{\infty} df P(f) \delta(\omega - \omega_{new} (\omega_1, f)) .
\end{equation}
The upper limit of the first $\omega_1$ integral can be taken as $\omega$ using our knowledge that $\omega_{new} (\omega_1, f)  > \omega_1$ and consequently when $\omega_1  > \omega$,
$\omega_{new} (\omega_1, f)$ must obey the inequality $\omega_{new} (\omega_1, f)  > \omega$ and therefore cannot contribute to the $\delta(\omega - \omega_{new} (\omega_1, f))$ integration term.

To perform this double integral let us first define a force $f_0$ by  
\begin{equation}
\label{eq7}
\omega_{new} (\omega_1, f_0) = \omega ,
\end{equation}
and then expand  
\begin{eqnarray}
\omega_{new} (\omega_1, f)  &\approx& \omega_{new} (\omega_1, f_0) + \frac{\partial \omega_{new}}{\partial f} |_{f=f_0} (f - f_0) \ , \nonumber\\
 &\approx& \omega + \frac{\partial \omega_{new}}{\partial f} |_{f=f_0} (f - f_0) \ .
\label{recon}
\end{eqnarray}
\noindent Using Eq.~(\ref{eq7}) we can now rewrite Eq.~(\ref{eq5}) as
\begin{equation}
\label{almost}
D(\omega ) = \int_0^\omega g_1(\omega_1) d\omega_1  \int_{-\infty}^{\infty} df P(f) \delta (\frac{\partial \omega_{new}}{\partial f} |_{f=f_0} (f - f_0)) .
\end{equation}
We can perform this integral exactly and find
\begin{equation}
\label{right}
D(\omega ) = \int_0^\omega g_1(\omega_1) d\omega_1 \frac{ P(f_0(\omega, \omega_1))}{ |\partial \omega_{new}/\partial f |_{f=f_0}|}.
\end{equation}
Combining Eq.~(\ref{eq7}) and Eq.~(\ref{omeganew}) we then find an explicit form for $f_0(\omega_1, \omega )$, namely
\begin{equation}
\label{f0depends}
f_0(\omega_1, \omega ) = \sqrt{(\omega^2 - \omega_1^2) (\frac{M^3 \omega_1^4}{3A})}.
\end{equation}
Recall that this is the force that contributes maximally to the DOS. Note also that over the whole range of the $\omega_1$ integration $(\omega^2 - \omega_1^2) >0$ and $f_0(\omega_1, \omega )$ given by Eq.~(\ref{f0depends}) is real.

We can also calculate $|\partial \omega_{new}/\partial f|^{-1}$ from Eq.~(\ref{omeganew}) and find
\begin{equation}
\label{wow}
|\partial \omega_{new}/\partial f |^{-1}= \big | \frac{M^3 \omega_1^4}{3 A f}\big | \big | \sqrt{\omega_1^2 + \frac{3 A f^2}{M^3 \omega_1^4}}\big |.
\end{equation}
To complete the calculation we need the functional form of $P(f)$ that is so far not specified. 
\subsection{The functional form of $P(f)$}
\label{Poff}

The distribution that we are seeking pertains to the forces that arise due to the interaction between our oscillators, displacing the low-frequency QLMs from their equilibrium positions by amounts $x_{i0}$. As a consequence they create random quenched static strains $f_i$ on each QLM due to similar displacements $x_{j0}$ of the other oscillators. Thus the force $f_i$ exerted on the $i$th
oscillator by the other $j$ oscillators  is of the form Eq.~(\ref{fi}).
Note we have chosen to treat  the displacements and forces here as scalars for simplicity. More accurately  we should treat both the displacements $x_{i0}^\alpha$ and the forces $f_i^{\alpha}$ as  components of a Cartesian vector and the coupling $J_{ij}^{\alpha\beta}$ as a second order tensor. Thus we would have $f_i^\alpha = \sum_j J_{ij}^{\alpha\beta} x_{j0}^\beta$. But here for simplicity, we consider these forces to be scalars with some distribution $P(f)$ .

Let us now consider $J_{ij}$ in more detail. We will assume there exists a spatial Poisson distributed  placement of QLMs at positions ${\bf r}_i$. In that case we can write
\begin{equation}
\label{eq8}
J_{ij} = y_{ij}/r_{ij}^\alpha.
\end{equation}
The random variable $y_{ij}$ will take care of the relative orientation of the QLMs and have zero mean. While $| {\bf r}_i -{\bf r}_j|$ is the distance between the QLMs. The exponent $\alpha$ depends on the nature of the interaction between the QLMs. For example the $\alpha = d$ for strain induced forces between the QLMs or electrostatic dipole-dipole interactions in the case of charged granular media. But other exponents may be important depending on the nature of the granular medium. There may exist charge-charge interactions in which case $\alpha = d-2$ may be a more suitable exponent. Or in the case case where higher order multipole interactions occur $\alpha > d$ may be important. Thus in general we may write
\begin{equation}
\label{eq9}
f_i = \sum_j J_{ij}x_{j0} = \sum_j \frac{y_{ij} x_{j0}}{r_{ij}^\alpha }.
\end{equation}
If we now examine Eq.~(\ref{eq9}) together with Eq.~(\ref{shift}), we see that we have a many-body nonlinear problem to solve, which may be suitable for simulations (as done in Sec.~\ref{GPS}) but well beyond analytical approach. {\em We therefore make an unavoidable uncontrolled approximation and replace Eq.~(\ref{eq9}) by the one body problem}
\begin{equation}
\label{eq10}
f_i = \sum_j J_{ij}x_{j0} \approx  \sum_j \frac{z_{ij}}{r_{ij}^\alpha }
\end{equation}
where the $z_{ij}$ are random variables with a given $P^*(z)$ of zero mean $\langle z\rangle_z =0$ and a given variance $\langle z^2 \rangle_z = \sigma_z^2$. We reiterate that this step, in addition to the assumption of the existence of the integral in Eq.~(\ref{C}) is not
guaranteed to apply to any realistic amorphous solid, and it needs to be assessed carefully in each case.

We could expect that $\sigma_z^2 = \langle (y_{ij}x_{i0})^2\rangle_{y x}$ could be estimated from simulations, but it is much better to treat the fluctuations $\sigma_z$ as a free parameter in the theory, and study how universal the predictions of the theory are with respect to changes in $\sigma_z$. In fact we will see that provided the fluctuations are bounded the exact value will not be too crucial.

We also need to stress that $P(f)$ does not follow the central limit theorem as $N \rightarrow \infty$. As mentioned in Ref. \cite{03GPS}, it does present strong similarities to the problem studied by Holtsmark and Chandrasekar when studying the gravitational force fluctuations in a system of $N$ galaxies. We shall therefore calculate $P(f)$ as follows. In the thermodynamic limit, all QLMs $f_i$ will have similar statistical properties. 
Let us now focus on one site $i$,  which we place at the origin of our coordinate system with $f_i=f = \sum_j z_j/r_j^\alpha$. Then
\begin{equation}
\label{eq11}
P(f)  =   \langle \delta (f -  \sum_j z_j/r_j^\alpha )\rangle = \frac{1}{2 \pi} \int_{- \infty}^{+ \infty}d\tau  e^{i f \tau }F(\tau) ,
\end{equation}
where $F(\tau )$ is the characteristic function associated with $P(f)$, namely

\begin{eqnarray}
\label{eq12}
F(\tau) &&=   \langle \exp\{-i \tau \sum_j z_j/r_j^\alpha \} \rangle  \\
&&=  (1/V^N)\int d{\bf r}_1 \cdots d{\bf r}_N \int dz_1 \cdots dz_N \nonumber \\ && \times P^*(z_1)\cdots P^*(z_N) \exp\{-i \tau \sum_j z_j/r_j^\alpha \} \nonumber \\
&& =  [(1/V)\int d{\bf r}  \int dz P^*(z) \exp\{-i \tau  z/r^\alpha \}]^N \nonumber \\
&& =  [1-(1/V)\int d{\bf r}  \int dz P^*(z) (1- \exp\{-i \tau z/r^\alpha  \})]^N \nonumber.
\end{eqnarray}
Eq.~(\ref{eq12})  can be written in this way as all the spatial integrations $\int d{\bf r}_1 \cdots d{\bf r}_N$ and all the random variable integrations $ \int dz_1 \cdots dz_N P^*(z_1)\cdots P^*(z_N) $ are independent of one another. Finally in the thermodynamic limit as $V \rightarrow \infty$, $N\rightarrow \infty$ with $n= N/V$ finite we find
\begin{eqnarray}
\label{eq13}
&&F(\tau)  =   [1-(1/V)\int d{\bf r}  \int dz P^*(z) (1- \exp\{-i \tau  z/r^\alpha  \})]^N \nonumber \\
&& \rightarrow  \exp{[ -  n \int d{\bf r}  \int_{-\infty}^{\infty}dz P^*(z) (1- \exp\{-i \tau z/r^\alpha  \})]} .
\end{eqnarray}

Performing the spatial integration observe every ${\bf r}$ can be matched by an equivalent ${\bf -r}$ contribution, and assuming $P^*(-z)= P^*(z)$ the integral becomes
\begin{equation}
\label{eq14}
F(\tau)  =  \exp{[ -  n 2 S_d \int_0^{\infty}\!\!\! r^{d-1}dr \!\! \int_0^{\infty}\!\!\! dz P^*(z) (1- \cos{\tau z/r^\alpha ) }]} .
\end{equation}
where $S_d$ is the surface area of a unit sphere in $d$ dimensions (i.e. $S_2 =2 \pi$ and $S_3 =4 \pi$ etc). We also note that $F(-\tau) = F(\tau)$ and that in consequence Eq.~(\ref{eq13}) is best integrated by introducing the new variable $y =  |\tau | | z|/r^\alpha$. Then we find
\begin{equation}
\label{eq15}
F(\tau)  =  \exp{[ -  \frac{ S_d n |\tau |^{d/\alpha }\langle |z|^{d/\alpha} \rangle_z}{\alpha } \int_0^{\infty} dy \frac{(1- \cos{y })}{y^{d/\alpha+1}} ]},
\end{equation}
where $\langle |z|^{d/\alpha} \rangle_z = 2  \int_0^{\infty} dz P^*(z) z^{d/\alpha}$. 

Let us analyze Eq.~(\ref{eq15}). We note that if $P^*(z)$ does not have a finite variance then $\langle |z|^{d/\alpha} \rangle_z$ diverges for $\alpha < d/2$. For example for charge-charge interactions in a charged amorphous solid. Further, even if a finite variance does exist, we note that the integral  $\int_0^{\infty} dy \frac{(1- \cos{y })}{y^{d/\alpha+1}}$ diverges for 
\begin{equation}
\label{eq16}
\alpha \le d/2.
\end{equation}
Now for charge charge interactions $\alpha = d-2$. For dimensions $d<4$, if charge-charge interaction plays a role in charged amorphous media, they may not possess a DOS obeying $D(\omega )\sim \omega^4$ at low frequencies. This should be valid for both $d=2$ and $d=3$. If, on
the other hand, it turns out that charged media do have $D(\omega )\sim \omega^4$, it is a strong indication that charge-charge
interactions are not playing an important role in the reconstruction of their density of states. 

Next use Eq.~(\ref{eq15}) to find the $P(f)$ in the case of elastic interactions. In this case $\alpha = d$ and Eq.~(\ref{eq15}) becomes
\begin{eqnarray}
\label{eq17}
F(\tau)  &=&  \exp{[ -  \frac{ S_d n |\tau | \langle |z| \rangle_z}{d } \int_0^{\infty} dy \frac{(1- \cos{y })}{y^2} ]}\ , \nonumber \\
&=&\exp{[ -  \frac{\pi S_d n |\tau | \langle |z| \rangle_z}{2d } ]}\ .
\end{eqnarray}
Substituting $F(\tau ) = \exp(-\delta f |\tau | )$ into Eq.~(\ref{eq11}) we find
\begin{eqnarray}
\label{eq18}
P(f) &  = &  \frac{1}{2 \pi} \int_{- \infty}^{+ \infty} d \tau e^{(i f-\delta f) |\tau| }   \\
& = &\frac{1}{\pi} \int_0^{\infty} d\tau \cos{f \tau } \exp(- |\delta f |\tau )  =  \frac{1}{\pi} \frac{|\delta f|}{(\delta f)^2 + f^2}\ . \nonumber
\end{eqnarray}
Thus we see we have a Lorentzian distribution with a  mean $\langle f \rangle $ and a standard deviation
\begin{equation}
\label{sigmaf}
\sigma_f =  \delta f =  \frac{\pi S_d n \langle |z| \rangle_z}{2d }  .
\end{equation}

Now combining Eqs. ~(\ref{eq1}), ~(\ref{almost}),~(\ref{right}),~(\ref{f0depends}) and ~(\ref{wow}) we find
\begin{eqnarray}
\label{loong}
&&D(\omega ) \approx  C \int_0^{\omega}\omega_1 d\omega_1 \frac{( \delta f/\pi )}{f_0(\omega_1, \omega )^2 +\delta f^2} \frac{M^3 \omega_1^4}{3 A f_0(\omega_1, \omega )} \nonumber \\&& \times\sqrt{\omega_1^2 + \frac{3 A f_0(\omega_1, \omega )^2}{M^3 \omega_1^4}} \nonumber \\
&& =  C \int_0^{\omega}\omega_1 d\omega_1 \frac{( \delta f/\pi )}{f_0(\omega_1, \omega )^2 +\delta f^2} \frac{M^3 \omega_1^4}{3 A } \nonumber \\ &&\times \sqrt{\frac{\omega_1^2}{f_0(\omega_1, \omega )^2 }+ \frac{3 A }{M^3 \omega_1^4}} \ .
\end{eqnarray}
Let us now introduce a new integration  variable $t$ by $\omega_1=\omega t$ then first from Eq.~(\ref{eq10})
\begin{equation}
\label{f0}
f_0(t, \omega ) = \omega^3 \sqrt{(1 - t^2) (\frac{M^3 t^4}{3A})}.
\end{equation}
and therefore
\begin{eqnarray}
\label{semifinal}
&&D(\omega )  =  \omega^4 C \int_0^{1}t dt \frac{( \delta f/\pi )}{ \omega^6 (1 - t^2) (\frac{M^3 t^4}{3A}) +\delta f^2} \frac{M^3 t^4}{3 A } \nonumber\\ && \times \sqrt{t^2/[(1 - t^2) (\frac{M^3 t^4}{3A})]+ \frac{3 A }{M^3 t^4}}
\end{eqnarray}
As $\omega \rightarrow 0$ this integral reduces to
\begin{equation}
\label{finalD}
D(\omega )  =  \frac{C \omega^4}{\pi \delta f} \int_0^{1}t^5 dt \frac{M^3}{3 A } \sqrt{t^2/[(1 - t^2) (\frac{M^3 t^4}{3A})]+ \frac{3 A }{M^3 t^4}}
\end{equation}
Eq.~(\ref{finalD}) yields the desired $D(\omega) \sim \omega^4$ behavior for the DOS.

In summary, the derivation of the universal density of states rests on one crucial assumption and one
uncontrolled approximation, as explained above. To test the crucial approximation Eq.~(\ref{eq10}) we
turn now to the GPS numerical model and examine it for different laws of interaction.
\section{The GPS model and numerical results}
\label{GPS}

In this section we explore further the model proposed by Gurevich, Parshin and Schober \cite{03GPS} which we 
denote as the GPS model. This model considers $N$ anharmonic oscillators on a three-dimensional lattice.
The $i$th oscillator is attached to the position $\B r_i$, and the total energy of the system is
\begin{equation}
U =\sum_i \left(\frac{1}{2} k_i x_i^2 +A x_i^4 \right) +\sum_{i\ne j} J_{ij} x_i x_j 
\end{equation} 
where $k_i$ are chosen randomly such that in the notation of Sec.~\ref{first} $g_0(\omega)\propto \omega^2$.
Note that with this choice one is guarantees (in three dimensions) the convergence of the integral (\ref{C}),
and see blow for a counter example. 
The coefficient of anharmonicity is chosen $A=1$. The interaction terms $J_{ij}$ are
\begin{equation}
J_{ij} \equiv J_0\frac{b_{ij}}{r_{ij}^\alpha}\ ,
\end{equation}
with $J_0=0.1$, $b_{ij}$ chosen from a flat distribution $b_{ij} \in [-0.5,0.5]$ and $\alpha$ controls
the range of interaction. We will explore below the values $\alpha=1,2$ and 3. The latter value
$\alpha=3$ is the one studied in \cite{03GPS}, resulting in $\omega^4$ law at small frequencies. 

Starting with the three-dimensional model with periodic boundary conditions, at each lattice site
we put an oscillator with an initial displacement $x_i$ taken randomly from the a uniform
distribution $x_i \in [-0.005,0.005]$. We use conjugate gradient minimization to obtain an
equilibrated configuration where the many-body problem of determining $x_{i0}$ is solved 
numerically. After computing the Hessian the eigenvalues and frequencies of the modes are found. Repeating the 
procedure with many random realizations, the density of states is determined by straightforward
binning. 

\begin{figure}
	\includegraphics[width=0.35\textwidth]{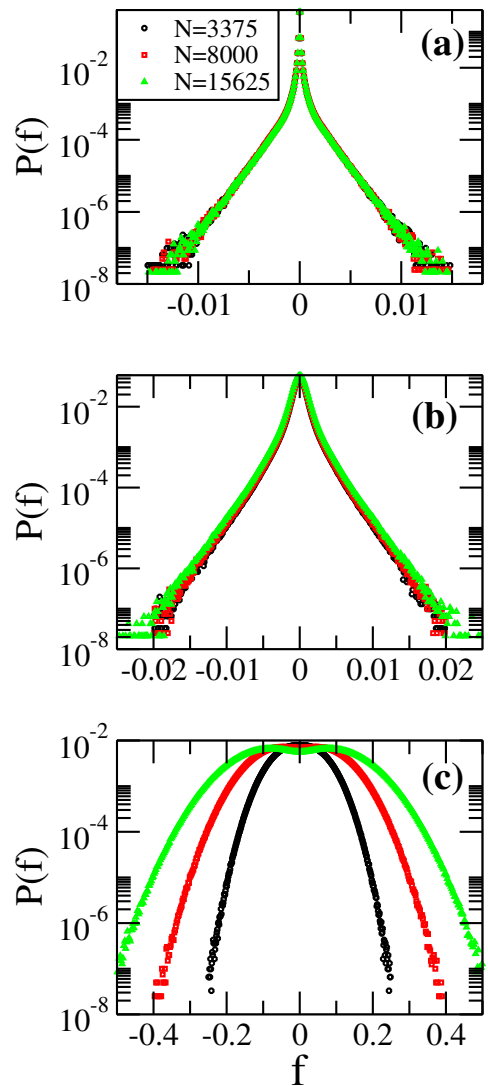}
	\caption{Probability distribution function of forces $P(f)$ for different system sizes, for the GPS model in
	three dimensions with $\alpha =3$ (panel a), $\alpha =2$ (panel b) and $\alpha =1$ (panel c). In all cases
	we employed 10000 independent realizations.
As expected from the analysis in Subsec.~\ref{Poff}, full convergence is observed in panel a, marginal 
convergence in panel b and no convergence in panel c.}
	\label{poffGPS}
	\end{figure}

As explained in Subsec~\ref{Poff} after Eq.~(\ref{eq9}), it is quite impossible to determine
analytically the distribution of forces $f_i$. But here we can do this easily, and in Fig.~\ref{poffGPS}
we show the probability distribution function (PDF) of
\begin{eqnarray}
f_i\equiv \sum_j J_{ij} x_{j0} \ ,
\end{eqnarray}
as a function of system size. Here $N=3375, 8000$ and 15625.  We expect from the analysis of Subsec.~\ref{Poff} that in dimension $d=3$ the PDF of forces will converge
nicely for $\alpha=3$, will be marginal for $\alpha=2$ and will not converge for $\alpha=1$. This is
precisely what we find, cf. Fig.~\ref{poffGPS}.
We note that for $\alpha=3$ not only that the PDF converges very well as a function of system size, it is very
close in form to a Lorentzian PDF as is expected by the theory. In contrast, for $\alpha=1$ not only that the
PDF of the forces does not converge as a function of the system size, it deviates more and more strongly
from a Lorentzian form when the system size increases. It even develops a dip at $f=0$, and if this dip
continues to develop for larger systems (outside the scope of our numerics at this point in time), we would
expect that the density of states with $\alpha=1$ would not follow the universal law (\ref{dof}).

\begin{figure}
	\vskip 0.8 cm
	\includegraphics[width=0.35\textwidth]{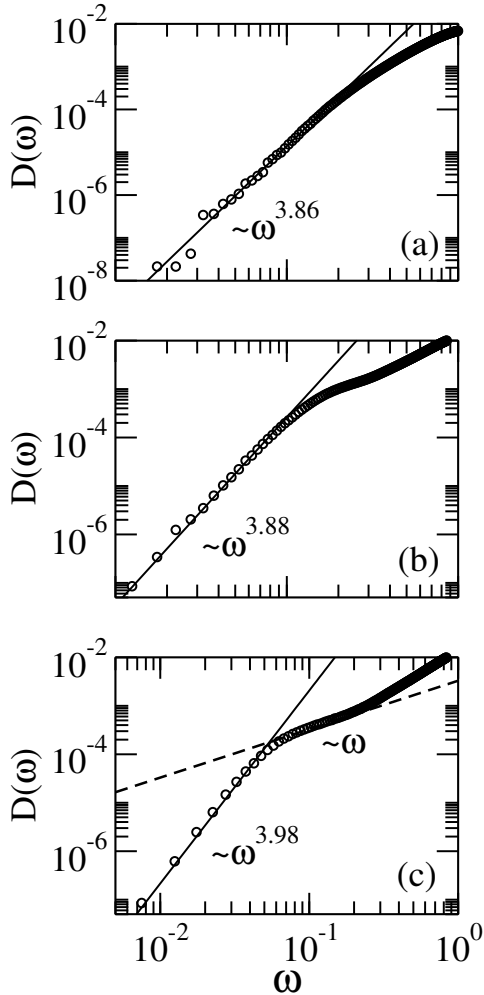}
	\caption{The low frequency regime of the density of states of the GPS model as found from 
		the largest available system size for $\alpha=1$ (panel a),  $\alpha=2$ (panel b) and
		 $\alpha=3$ (panel c).  In all cases
		 we employed 10000 independent realizations. The continuous lines are best linear fits to the data. The dashed line
	 in panel c has a slope of unity, revealing the regime of the first reconstruction, in agreement with Eq.~(\ref{eq1}).}
\label{dosGPS}
\end{figure}

Contrary to this expectation, the direct measurement of the density of states does not show a major
difference at the low frequency regime for the different values of $\alpha$. This can be seen in Fig.~\ref{dosGPS}
where results are shown for the largest system size available (i.e. $N$=15625) for the three values of $\alpha$.
One cannot say that there is a very large difference between the resulting scaling laws. This is an indication that
the actual form of a Lorentzian PDF that is employed in Subsec.~\ref{Poff} is not really needed,
and it is sufficient in fact that $P(f=0)$ is not zero. Whether or not for $\alpha=1$
$P(f=0)\to 0$ when $N\to \infty$ is a question that cannot be answered at present and has
to remain for future analysis. 

\begin{figure}
	\vskip 0.8 cm
	\includegraphics[width=0.35\textwidth]{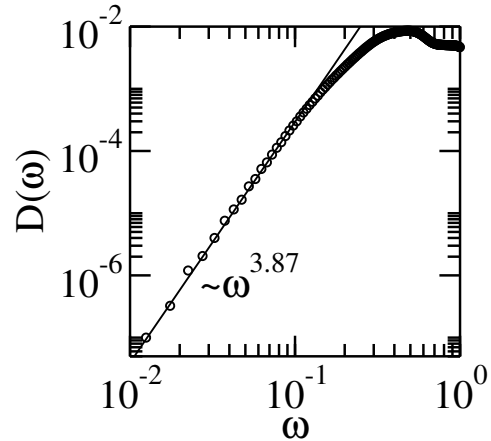}
	\caption{The low frequency regime of the density of states of the GPS model computed with 
		a uniform distribution $g_0$ in the interval [0,1]. Here we employ 5000 realizations of a system
		with $N=$ 8000. We see a loss of the linear
		regime $g_1\propto \omega$, as expected, but we still get the universal behavior at low
		frequencies!}
	\label{bust}
	\end{figure}

It is relevant however to comment that for $\alpha=3$ we can observe in Fig.~\ref{dosGPS}
the result of the {\em first} reconstruction. This is evidenced by the slope of about unity
as indicated by the dashed line, which is in agreement with $g_1(\omega) \propto \omega$ at intermediate low
frequencies. It is therefore interesting to destroy by hand the convergence of the integral 
in Eq.~(\ref{C}). We do it by choosing $g_0$ to be uniform in the interval [0,1]. 
we expect to lose the linear dependence of $g_1(\omega)$, with unknown consequences for the final
form of $D(\omega)$. Indeed, in Fig.~\ref{bust} we present the density of states for this choice
of the GPS model with $\alpha=3$. The result is quite surprising: the linear regime with a slope unity disappears, 
but the low frequencies of the density of states still conform (to as good an approximation as above)
with Eq.~(\ref{eq1}). The lesson drawn from this and the other numerical results shown above is provided in 
Sec.~\ref{summary}.
\section{Summary and Discussion}
\label{summary}

In summary, we first examined numerically the density of states of a model of charged granular
solid, and found that in the lowest frequency end it agrees with the universal law Eq.~(\ref{dof}).
Since this model contains long range interactions, we returned to the theoretical considerations of
Refs.~\cite{91BGGS,03GC,03GPS,07PSG} to clarify the expectations of how the density of states should
depend on different laws of interactions. We reviewed this theory paying attention to the assumptions and
approximation made. We found that the theory predicts a failure in the first reconstruction when $\alpha$ in
Eq.~(\ref{eq8})  is too small. We also expect a failure in the second reconstruction in such a case,
	since the PDF $P(f)$ is not expected to converge. Direct calculations using the GPS model verify
	that indeed $P(f)$ does not converge for $\alpha=1$ and $d=3$, and the linear reconstruction also
	disappears. And yet, the direct calculation of the density of states showed that Eq.~(\ref{dof}) is 
	extremely robust, oblivious of all these delicacies. This finding indicates strongly that further
	research to uncover the surprising robustness of this scaling law is still called for. It is not
	impossible that one reason for this robustness is that the model does not suffer from the approximation
	embodied in the transition from Eq.~(\ref{eq9}) to Eq.(\ref{eq10}). The divergence that the theory
	predicts may result from this approximation. We hope that this
	paper will inspire future research to unravel further the interesting issues discussed here. 
	
	\acknowledgments
	We thank Eran Bouchbinder for bringing the GPS model to our attention, including the role of the exponent $\alpha$. This work has been supported in part by the US-Israel Binational Foundation and by the cooperation project
	COMPAMP/DISORDER jointly funded by the Ministry of Foreign Affairs and International Cooperation(MAECI)
	of Italy and by the Ministry of Science and Technology (MOST) of Israel.
	E.~L.~acknowledges support from the NWO (Vidi grant no.~680-47-554/3259)

\bibliography{biblio}

\end{document}